# Chemical tuning of photo- and persistent luminescence of $Cr^{3+}$-activated β-$Ga_2O_3$ by alloying with $Al_2O_3$ and $In_2O_3$

Vasyl Stasiv[a,*], Yaroslav Zhydachevskyy[a,b], Vitalii Stadnik[c], Vasyl Hreb[c], Vitaliy Mykhaylyk[d], Leonid Vasylechko[c], Adriy Luchechko[e], Tomasz Wojciechowski[a], Piotr Sybilski[a], Andrzej Suchocki[a]

[a] *Institute of Physics, Polish Academy of Sciences, aleja Lotników 32/46, Warsaw 02-668, Poland*

[b] *Berdyansk State Pedagogical University, Shmidta Str. 4, Berdiansk 71100, Ukraine*

[c] *Lviv Polytechnic National University, S. Bandera Str. 12, Lviv 79013, Ukraine*

[d] *Diamond Light Source, Harwell Campus, Didcot, OX11 0DE, UK*

[e] *Ivan Franko National University of Lviv, Tarnavskogo Str. 107, Lviv 79017, Ukraine*

*Email: stasiv@ifpan.edu.pl

**ABSTRACT**

An effect of alloying of the monoclinic β-$Ga_2O_3$ with $Al_2O_3$ and $In_2O_3$ on the photoluminescent, thermoluminescent and persistent luminescent properties of $Cr^{3+}$ ions has been comprehensively investigated. For this purpose, various series of $Cr^{3+}$ and $Ca^{2+}$ co-doped microcrystalline phosphors were synthesized by the solution combustion method, including pseudobinary compounds like (Ga-Al)$_2$O$_3$ with up to 20% Al and (Ga-In)$_2$O$_3$ with up to 50% In as well as pseudoternary compounds (Ga-Al-In)$_2$O$_3$ with balanced proportion of Al, Ga and In. The phase composition and crystal structure of the obtained materials were examined by X-ray powder diffraction technique. Detailed luminescence studies were conducted for the (Ga-Al)$_2$O$_3$ and (Ga-In)$_2$O$_3$ compounds which exhibited a single-phase monoclinic structure. Low-temperature and time-resolved photoluminescence investigations of the Cr-doped pseudobinary compounds unveiled several types of $Cr^{3+}$ centers, attributed to the Al-, Ga- and In-centered octahedra in the studied alloys. The obtained results underscore the benefit of bandgap engineering through alteration in the host lattice chemical composition for efficient tuning of the thermoluminescent and persistent luminescent properties of the near-infrared-emitting β-$Ga_2O_3$:Cr based phosphors. Furthermore, it was demonstrated that modification of the chemical composition of the host lattice also adjusts the thermometric performance of the studied phosphors. Indeed, the specific sensitivity of the β-Ga$_2$O$_3$:Cr$^{3+}$ decay time luminescence thermometer showed nearly twofold enhancement when the host lattice was alloyed with 30% of In$_2$O$_3$.

**INTRODUCTION**

Trivalent chromium (Cr$^{3+}$) and tetravalent manganese (Mn$^{4+}$) both possessing 3$d^3$ electronic configuration, stand out as widely recognized and frequently employed non-rare-earth activators capable of generating emission in the deep red and near-infrared (NIR) regions. The luminescent properties of these ions have been extensively explored since the early '60s, marked by a groundbreaking demonstration of laser emission from Cr$^{3+}$-doped α-Al$_2$O$_3$ (ruby) by Maiman.[1] Following this, a plethora of crystalline compounds, particularly oxides, doped with Cr$^{3+}$ and Mn$^{4+}$ ions have been investigated (see *e.g.* Refs. [2–5] and references therein), and their potential has been demonstrated across a wide spectrum of applications encompassing tunable lasers, optical converters for white lightening, pressure and temperature sensors, NIR light sources and NIR persistent phosphors for biological imaging, food freshness analysis, agriculture, human health monitoring, etc.[6–13]

One of the most promising and extensively researched phosphors activated with Cr$^{3+}$ ions is β-Ga$_2$O$_3$. Gallium oxide is a well-known wide-bandgap semiconductor with diverse applications spanning semiconductor electronics, optoelectronics, ultraviolet (UV) photodetectors, and more.[14–19] Ga$_2$O$_3$ exhibits several structural polymorphs including the corundum-like (α), monoclinic (β), defective spinel (γ), and two variations (ε and δ) of orthorhombic structure.[20] Among these, the monoclinic phase (β-Ga$_2$O$_3$) stands out as the most prevalent being both the most stable and readily attainable under ambient conditions.[21–24] When activated with Cr$^{3+}$ ions, the β-Ga$_2$O$_3$ produces a broad and efficient emission in the wavelength range extending from about 650 to 950 nm at room temperature.[4,25–27] It is noteworthy that Cr$^{3+}$ ions in β-Ga$_2$O$_3$ are situated in octahedral coordination experiencing intermediate crystal field strength that results in both the sharp line (the spin-forbidden $^2$E → $^4$A$_2$ transition) and broadband (the spin-allowed $^4$T$_2$ → $^4$A$_2$ transition) emissions, that corresponds to O-Cr-B type emission according to classification of Adachi[4]. The distinctive emission features of β-Ga$_2$O$_3$:Cr$^{3+}$ phosphor hold significant potential for various applications such as NIR tunable lasers,[7] artificial solid-state lighting for agriculture,[28] sensors for high-pressure calibration of diamond anvil cell (DAC),[29] or non-contact luminescence thermometry.[30,31] Recent studies have also demonstrated that $\beta$-Ga$_2$O$_3$:Cr$^{3+}$ co-doped with Ca$^{2+}$ is a promising NIR persistent luminescence phosphor with enhanced properties.[32]

Alloying of gallium oxide with aluminum, scandium or indium oxide to create solid solutions is a well-known method for tuning their physical properties over a wide range. This is primarily achieved by modifying the crystal structure and energy band gap of the material. For instance, substituting about 80% of Ga with Al in $\beta$-Ga$_2$O$_3$ leads to the transformation from the ambient-pressure-synthesized material to the rhombohedral ($\alpha$-type) structure.[33–36] Conversely, replacing Ga with larger In atoms can result in a transformation towards the In$_2$O$_3$-type structure when In amount exceeds 41%[37] or 50%.[38]

Even in the confines of a monoclinic-type structure crystal parameters of $\beta$-Ga$_2$O$_3$ can be gradually adjusted by partially replacing Ga with Al or Sc (In).[33–35,37–41] Moreover the electronic bandgap of gallium oxide can be expanded by alloying with aluminum oxide, or conversely reduced by alloying with scandium or indium oxide.[33,39,40,42] The simultaneous alloying of $\beta$-Ga$_2$O$_3$ with Al and In oxides, leading to the formation of quaternary (Al$_x$In$_y$Ga$_{1-x-y}$)$_2$O$_3$ alloys, holds significant interest as it potentially opens up even greater avenues for tailoring the material physical properties. However, up to now, this possibility has only been explored in theoretical studies.[43]

When Cr$^{3+}$ ions are introduced into the material, the photoluminescent properties of these ions can be tuned significantly. This is primarily achieved by altering the crystal field strength experienced by the activator, either through chemical pressure (CP) by adjusting the chemical composition of the host[40–42,44] or by applying the hydrostatic pressure.[34,41] Notably, in the case of Al$_2$O$_3$-Ga$_2$O$_3$ solid solutions substituting Al with Ga weakens the crystal field strength experienced by Cr$^{3+}$ ions. This leads to a gradual increase in the fraction of the broadband emission from Cr$^{3+}$ with respect to the narrow-line emission.[34,42,44] Furthermore, replacing Ga with Sc[41] or In[40] can further weaken the crystal field strength experienced by Cr$^{3+}$ ions. This results in the complete extinction of the narrow-line emission at room temperature and a shift of the broadband emission towards the infrared region.

Recently we demonstrated that Mn$^{4+}$ ions in the $\beta$-Al$_2$O$_3$-Ga$_2$O$_3$ solid solutions form two types of distinct centers, denoted as Mn$_\text{I}$($\beta$) and Mn$_\text{II}$($\beta$), which were tentatively attributed to Al and Ga octahedral sites in $\beta$-Al$_2$O$_3$-Ga$_2$O$_3$, respectively.[35] Solid solutions containing simultaneously Al, Ga and In atoms have not been experimentally examined thus far.

The non-single exponential decay of Cr$^{3+}$ photoluminescence reported in $\beta$-Ga$_2$O$_3$-In$_2$O$_3$[40] can also suggest the possibility of Cr$^{3+}$ multicenters in such alloys. However, to the best of our knowledge, no spectroscopic features indicating the presence of $Cr^{3+}$ multicenters have been identified in any of the $(Al/Ga/Sc/In)_2O_3$ solid solutions to date.

In order to shed light on the nature of the $Cr^{3+}$ centers and investigate in more detail the effects of the crystal field strength on their photoluminescence within such solid solutions we instigated a comprehensive investigation. We synthesized a series of microcrystalline powders with nominal compositions $(Ga_{1-x}In_x)_2O_3$:Cr(0.05at.%), Ca(0.5at.%) ($x$ = 0.05; 0.1; 0.15; 0.2; 0.3; 0.4; 0.5), $(Ga_{1-y}Al_y)_2O_3$:Cr(0.05at.%), Ca(0.5at.%) ($y$ = 0; 0.1; 0.2), $(Ga_{0.25}Al_{0.5}In_{0.25})_2O_3$:Cr(0.05at.%), Ca(0.5at.%), and $(Ga_{0.33}Al_{0.34}In_{0.33})_2O_3$: Cr(0.05at.%), Ca(0.5at.%) and systematically studied by powder X-ray diffraction (XRD), scanning electron microscopy (SEM) and luminescence techniques. Co-doping with $Ca^{2+}$ was used here to enhance their energy storage capability and persistent luminescent (PersL) properties following the studies of Luchechko et al.[45] and Sun et al.[32] Specifically, we conducted in-depth study of the photoluminescence (PL) spectra, PL excitation spectra, and PL decay kinetics of the materials in the temperature range from 4 to 350 K. Additionally, the PersL decays and the thermoluminescence (TL) after UV excitation were measured in the temperature range from 77 to 600 K. Our studies were aimed also towards exploring the potential for fine-tuning the performance of the studied materials as NIR persistent phosphors and non-contact luminescence thermometers through alterations in the chemical composition of the β-$Ga_2O_3$-based host material.

**EXPERIMENTAL METHODS**

Two series of Cr and Ca co-doped gallium-indium and gallium-aluminum oxides of nominal compositions $(Ga_{1-x}In_x)_2O_3$:Cr(0.05at.%), Ca(0.5at.%) ($x$ =0; 0.05; 0.1; 0.15; 0.2; 0.3; 0.4; 0.5) and $(Ga_{1-y}Al_y)_2O_3$:Cr(0.05at.%), Ca(0.5at.%) ($y$ = 0.1; 0.2) were synthesized by solution combustion method using the mixture of glicine and urea as fuel from $Al(NO_3)_3·9H_2O$, $Cr(NO_3)_3·9H_2O$ and $Ca(NO_3)_2·4H_2O$ as initial reagents. The metallic Ga and $In_2O_3$, dissolved in concentrated $HNO_3$ were used as gallium and indium sources, respectively. Appropriate aliquots of metal nitrate solutions corresponding to a nominal composition of a sample were mixed on a magnetic stirrer for 30 min. After that, the glycine and urea water solutions were added to the reaction mixture ensuring the molar ratios between metals, glycine, and urea were equal to 1.2:1.0:1.5. The prepared solutions were evaporated at the temperature of 373 K to minimize the total volume of the samples in the drying cabinet during 2-4 hours to the formation of a brownish gel-like solution. After that, the samples were moved into the 873 K muffle furnace for a few minutes to remove the organic components and formation of the desired phase. The final heat treatment of the products was performed at 1773 K for 5 h. The pseudoternary mixed gallium-indium-aluminium oxides of nominal compositions $(Ga_{0.25}Al_{0.5}In_{0.25})_2O_3$:Cr(0.05at.%), Ca(0.5at.%), and $(Ga_{0.33}Al_{0.34}In_{0.33})_2O_3$: Cr(0.05at.%), Ca(0.5at.%) were synthesized by using the same method.

X-ray powder diffraction (XRD) characterization of the synthesized materials was performed by an Aeris benchtop powder diffractometer (Malvern Panalytical) equipped with a PIXcel$^{1D}$ strip detector. Experimental diffraction data were collected using filtered Cu K$\alpha$ radiation ($\lambda$ = 1.54185 Å) in a 2$\theta$ range of 6 - 105 degrees with a 2$\theta$ step of 0.02º. Lattice parameters, coordinates and displacement parameters of atoms were derived from experimental XRD patterns by full profile Rietveld refinement using the WinCSD software package.[46]

The photoluminescence (PL) and the photoluminescence excitation (PLE) spectra were measured using a Horiba/Jobin-Yvon Fluorolog-3 spectrofluorometer with a 450 W continuous spectrum xenon lamp for excitation. The emission was detected by a Hamamatsu R928P photomultiplier operating in a photon counting mode. The PL spectra were corrected for the spectral response of the used system. The luminescence decay kinetics were measured using the same Fluorolog-3 spectrofluorometer with the excitation light modulated by a mechanical chopper. The spectroscopic measurements in the temperature range of 4.4-350 K were carried out in a Janis continuous-flow liquid helium cryostat using a Lake Shore 331 temperature controller. The studies in the temperature range of 77-600 K were done using a Linkam THMS600 temperature stage.

**RESULTS AND DISCUSSION**

**Phase composition and crystal structure parameters of the studied compounds**

The XRD analysis revealed that both series of the Al- and In-substituted samples adopt a monoclinic β-$Ga_2O_3$ type of structure, similar to the parent $Ga_2O_3$ material (Figure 1). The shift of diffraction maxima towards the right- and left-side in $(Ga_{1-y}Al_y)_2O_3$ and $(Ga_{1-x}In_x)_2O_3$ series respectively in comparison with $Ga_2O_3$ structure indicate the pertinent lattice contraction/expansion resulting from the partial substitution of $Ga^{3+}$ ions with either smaller $Al^{3+}$ or larger $In^{3+}$ species. The respective radii of the cations in the studied materials according to the Shannon [47] scale are given in Table 1. The peculiarity of the experimental XRD patterns of the In-containing samples is a detectable redistribution of the intensities of

Bragg's maxima (see Figure 1) due to a pronounced multiple texture (preferred orientation) of the powders being especially pronounced in [100] direction.

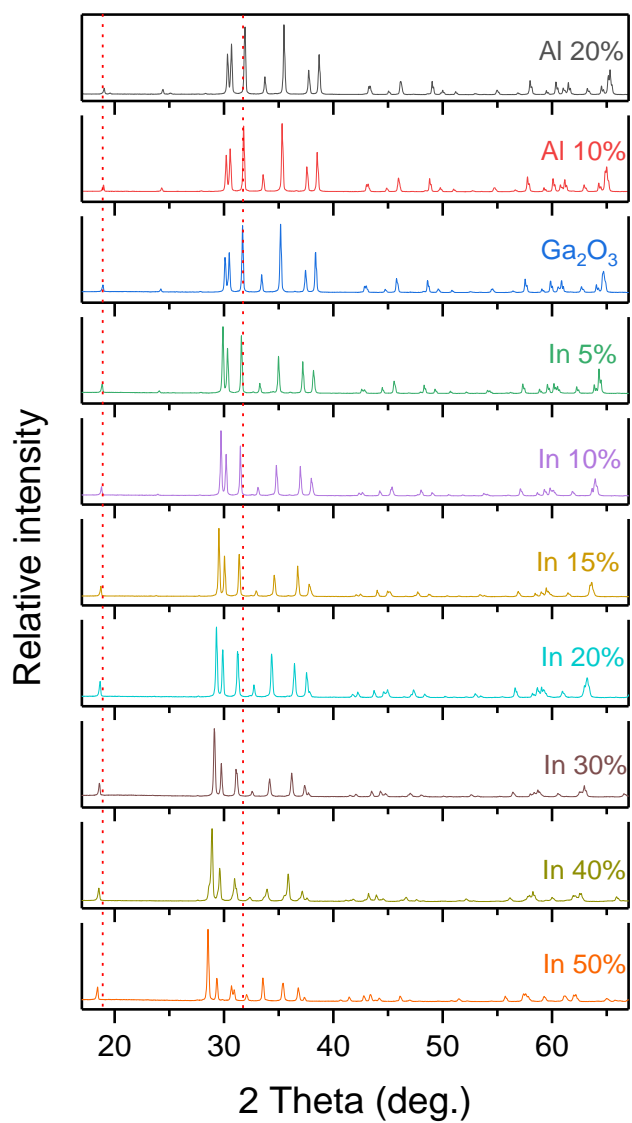

**Figure 1.** Experimental XRD patterns of $(Ga_{1-y}Al_y)_2O_3$ and $(Ga_{1-x}In_x)_2O_3$ series compared with $Ga_2O_3$:Cr, Ca sample.

**Table 1.** Ionic radii of cations (in Å) for coordination number (CN) 6 and 4 from Ref. [47]

| Ion | CN = 6 | CN = 4 |
| --- | --- | --- |
| $Ga^{3+}$ | 0.62 | 0.47 |
| $Al^{3+}$ | 0.54 | 0.39 |
| $In^{3+}$ | 0.80 | 0.62 |
| $Cr^{3+}$ | 0.615 | - |
| $Ca^{2+}$ | 1.00 | - |

Further analysis of structural peculiarities of the $(Ga_{1-y}Al_y)_2O_3$ and $(Ga_{1-x}In_x)_2O_3$ compounds was carried out by full profile Rietveld refinement performed in β-$Ga_2O_3$ structural model, space group $C2/m$. For the $(Ga_{1-y}Al_y)_2O_3$:Cr, Ca structures, the refinement procedure led to an excellent agreement between experimental and calculated XRD profiles (see example in Figure 2). The refinement of site occupancies revealed that $Al^{3+}$ ions occupy both tetrahedral and octahedral positions within the β-$Ga_2O_3$ structure but with a clear preference for the octahedral site. The refined $Al_{tet}$:$Al_{oct}$ ratio in the $(Ga_{1-y}Al_y)_2O_3$:Cr, Ca samples with y=0.1 and 0.2 was found to be 0.07:0.11 and 0.11:0.27, respectively. These values closely align with the 1:2 ratio reported for β-$Ga_{2-x}Al_xO_3$ (x=0.2-1.3) structures from Rietveld refinement and $^{27}Al$ NMR spectroscopy data.[48] With a progressive substitution of Al for Ga sites in β-$Ga_2O_3$ structure and a further increase of aluminum content in $(Ga_{1-y}Al_y)_2O_3$:Cr, Ca series a decrease in the lattice parameters and unit cell volume is evident (see Figure 2, pink squares). This effect is attributed to the smaller radius of $Al^{3+}$ cation in comparison with $Ga^{3+}$ one.

In the case of indium-containing $(Ga_{1-x}In_x)_2O_3$:Cr, Ca series, achieving satisfactory results of Rietveld refinement required including texture correction in the refinement procedure. Better results were achieved after the correction of texture in the [100] direction. An illustration of the Rietveld refinement results for the $(Ga_{0.8}In_{0.2})_2O_3$:Cr, Ca sample is presented in the bottom panel of Figure 2. The obtained structural data and corresponding residuals are detailed in Table 2. Further refinement of site occupancies revealed, that in contrast to the above described $(Ga_{1-y}Al_y)_2O_3$ structures, the $In^{3+}$ ions in $(Ga_{1-x}In_x)_2O_3$:Cr, Ca series occupy exclusively octahedral positions within the β-$Ga_2O_3$ structure, whereas the tetrahedral sites remain solely occupied with $Ga^{3+}$ ions (see Table 2 and insets in Figure 2). This observation correlates with the recent finding reported by Zhong et al. for the $Ga_{2-x}In_xO_3$:$Cr^{3+}$ series.[40]

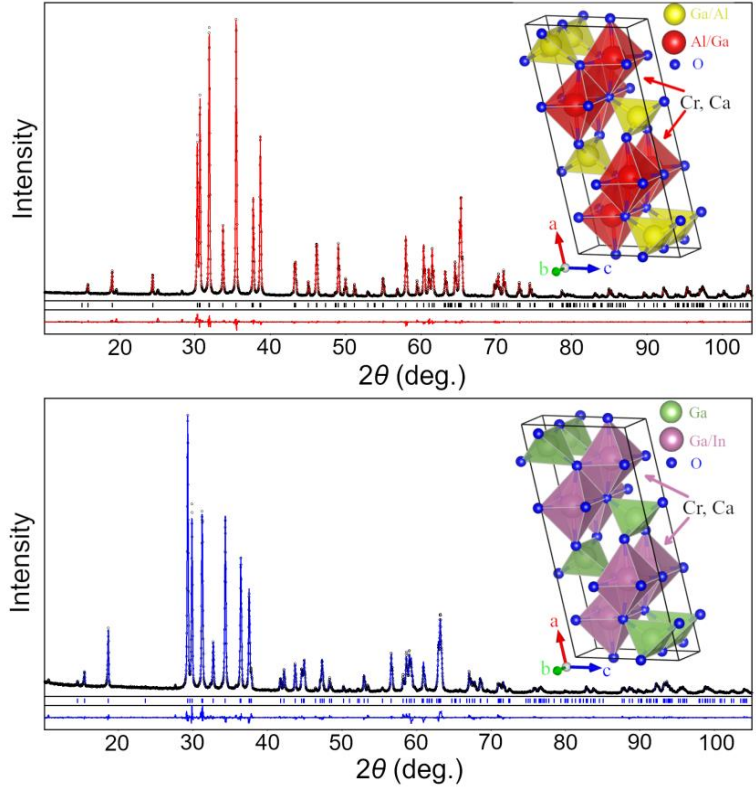

**Figure 2.** Graphical results of Rietveld refinement of $(Ga_{0.8}Al_{0.2})_2O_3$:Cr, Ca (top) and $(Ga_{0.8}In_{0.2})_2O_3$:Cr, Ca (bottom) structures. Experimental XRD patterns (small black circles) are shown in comparison with calculated patterns (red and blue lines, respectively). Insets show polyhedral representations of corresponding structures.

**Table 2.** Lattice parameters, coordinates, and displacement parameters of atoms in monoclinic structures of $(Ga_{0.8}Al_{0.2})_2O_3$:Cr, Ca and $(Ga_{0.8}In_{0.2})_2O_3$:Cr, Ca samples (SG $C2/m$, $Z=4$)

| Lattice parameters | Atoms, sites | $x/a$ | $y/b$ | $z/c$ | $B_{iso/eq}$, Å$^2$ | Occupancy |
|---|---|---|---|---|---|---|
| $(Ga_{0.8}Al_{0.2})_2O_3$:Cr, Ca; $R_I = 0.0264$, $R_P = 0.0512$ | | | | | | |
| $a$=12.1379(1) Å $b$=3.01341(4) Å $c$=5.77224(6) Å $\beta$=103.932(1) ° | Ga1, 4$i$ | 0.09023(9) | 0 | 0.7951(2) | 1.18(3) | 0.89(2) Ga$^{3+}$ + 0.11(2) Al$^{3+}$ |
| | Ga2, 4$i$ | 0.34162(8) | 0 | 0.6860(2) | 0.88(3) | 0.73(2) Ga$^{3+}$ + 0.27(2) Al$^{3+}$ |
| | O1, 4$i$ | 0.1614(3) | 0 | 0.1128(8) | 1.28(10) | O$^{2-}$ |
| | O2, 4$i$ | 0.4957(3) | 0 | 0.2558(6) | 2.17(12) | O$^{2-}$ |
| | O3, 4$i$ | 0.8292(3) | 0 | 0.4410(9) | 0.83(9) | O$^{2-}$ |
| $(Ga_{0.8}In_{0.2})_2O_3$:Cr, Ca; $R_I = 0.0448$, $R_P = 0.0800$ | | | | | | |
| $a$=12.5141(2) Å $b$=3.11373(7) Å $c$=5.8818(1) Å $\beta$=103.278(1) ° | Ga1, 4$i$ | 0.0901(1) | 0 | 0.7930(3) | 0.81(4) | 0.99(2) Ga$^{3+}$ |
| | Ga2, 4$i$ | 0.34391(8) | 0 | 0.6887(3) | 0.74(3) | 0.62(3) Ga$^{3+}$ + 0.38(3) In$^{3+}$ |
| | O1, 4$i$ | 0.1633(4) | 0 | 0.1169(12) | 0.52(15) | O$^{2-}$ |
| | O2, 4$i$ | 0.4858(4) | 0 | 0.2581(10) | 1.7(2) | O$^{2-}$ |
| | O3, 4$i$ | 0.8269(4) | 0 | 0.4397(14) | 1.6(2) | O$^{2-}$ |
| Texture axis and parameter: [1 0 0]   0.364(3) | | | | | | |

To understand the possible reason for the pronounced texture phenomenon in the In-containing series, target SEM examinations of the selected samples were performed (Figure 3). The analysis revealed that the Al-containing samples, much like the unsubstituted $Ga_2O_3$ sample, consist of relatively small (< 2 µm) well-shaped faceted grains, which are (almost) randomly distributed through the volume of the powders. In contrast, the In-containing samples exhibit several relatively large (> 20-50 µm) distinctively shaped and oriented crystallites, which are likely responsible for the observed redistribution of diffraction maxima intensities. Furthermore, it was observed that the texture effect increases with increasing indium content in $(Ga_{1-x}In_x)_2O_3$:Cr, Ca series, as supported by both Rietveld refinement and SEM data. It should be noted that no textures were observed in the $Ga_{2-x}In_xO_3$:$Cr^{3+}$ powders prepared by a solid-state chemical route at 1300 ºC by Zhong et al.[40] On a different note, the experimental XRD patterns of our $(Ga_{1-x}In_x)_2O_3$:Cr, Ca materials obtained by the solution combustion method closely resembles the oriented β-$Ga_2O_3$ films deposited on the sapphire (0001) substrates by the radio-frequency magnetron sputtering as demonstrated by Dong et al.[49] It is evident, that peculiarities of the solution combustion process used for the synthesis of $Ga_2O_3$-based materials in the present work favor the formation and rapid growth of the large crystallites in the case of In-containing materials.

The lattice parameters *a*, *b* and *c* exhibit a systematic increase with rising In content in the $(Ga_{1-x}In_x)_2O_3$:Cr, Ca series, whereas the monoclinic angle β experiences a decrease (Figure 4). This behavior aligns with the previous observations reported in Refs.[37,38,40,50]. However, upon closer examination of the concentration dependence of the unit cell dimensions in $(Ga_{1-x}In_x)_2O_3$:Cr, Ca series a notable deviation from the expected monotonic trend is discernable between *x*= 0.2 and 0.3 (see Figure 4). The lattice parameters and unit cell volumes of the samples with *x*=0.3, 0.4 and 0.5 are positioned clearly below the empirical linear fits established for the five data sets of the powder and single crystal samples of $(Ga_{1-x}In_x)_2O_3$.[39] This phenomenon prompts the question of whether it is a peculiarity of the $(Ga_{1-x}In_x)_2O_3$:Cr, Ca materials studied in our work, or it is a more general feature of the $Ga_2O_3$-$In_2O_3$ pseudobinary system. The fact, that the lattice parameters of the undoped material with *x*=0.5 reported by Shannon and Previtt[38], in their pioneering work on the $Ga_2O_3$-$In_2O_3$ system, also exhibit a similar deviation from the linearity as indicated by red stars in Figure 4, suggests that the observed phenomenon may be inherent for this mixed oxide system. In any case, a thorough examination of the phase and structural behavior in the β-

Ga$_2$O$_3$-In$_2$O$_3$ system using modern complementary experimental techniques is warranted to gain deeper insight into this issue.

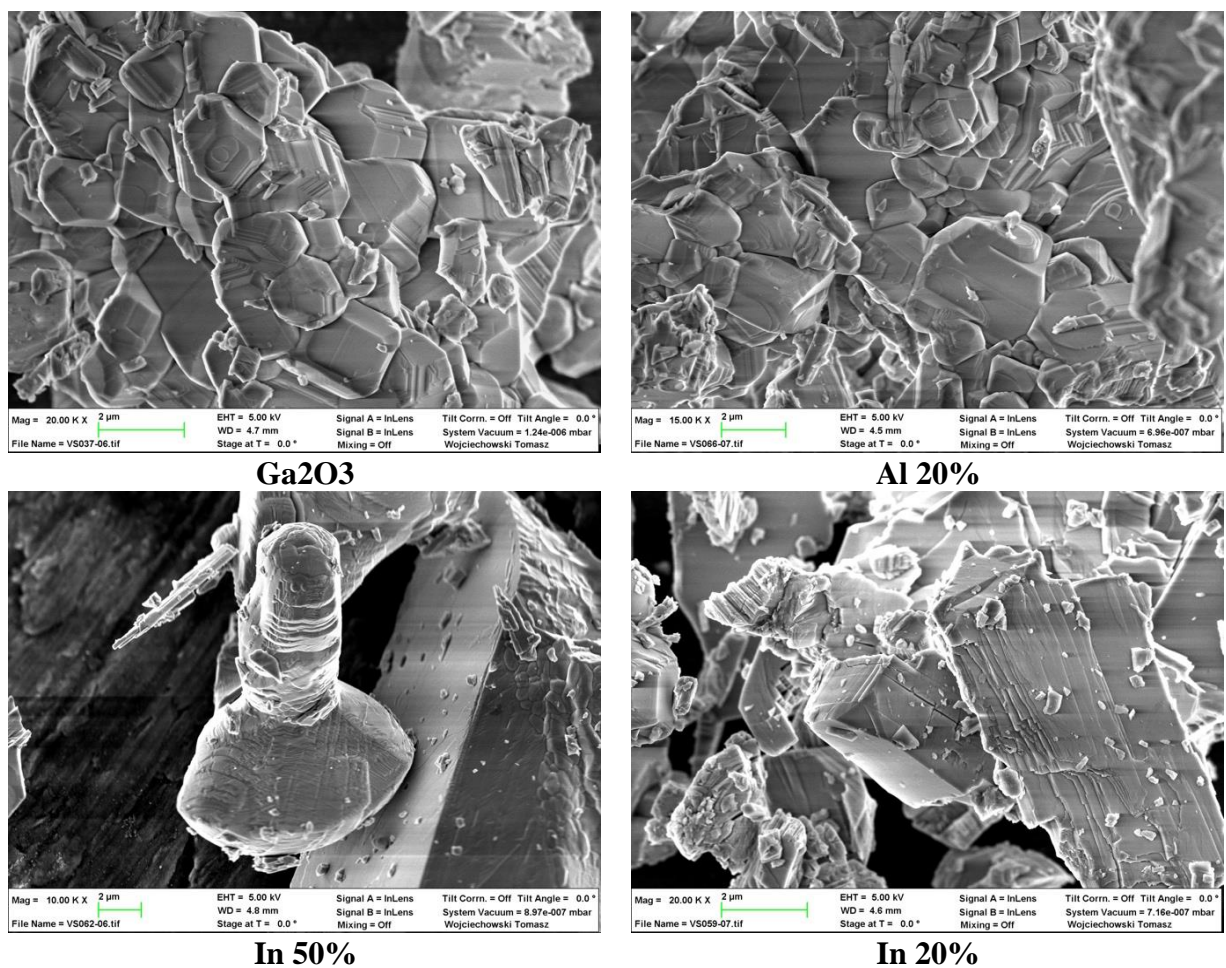

**Figure 3.** Selected SEM images of the Ga$_2$O$_3$:Cr, Ca, (Ga$_{0.8}$Al$_{0.2}$)$_2$O$_3$:Cr, Ca, (Ga$_{0.8}$In$_{0.2}$)$_2$O$_3$:Cr, Ca and (Ga$_{0.5}$In$_{0.5}$)$_2$O$_3$:Cr, Ca powders obtained by the solution combustion method.

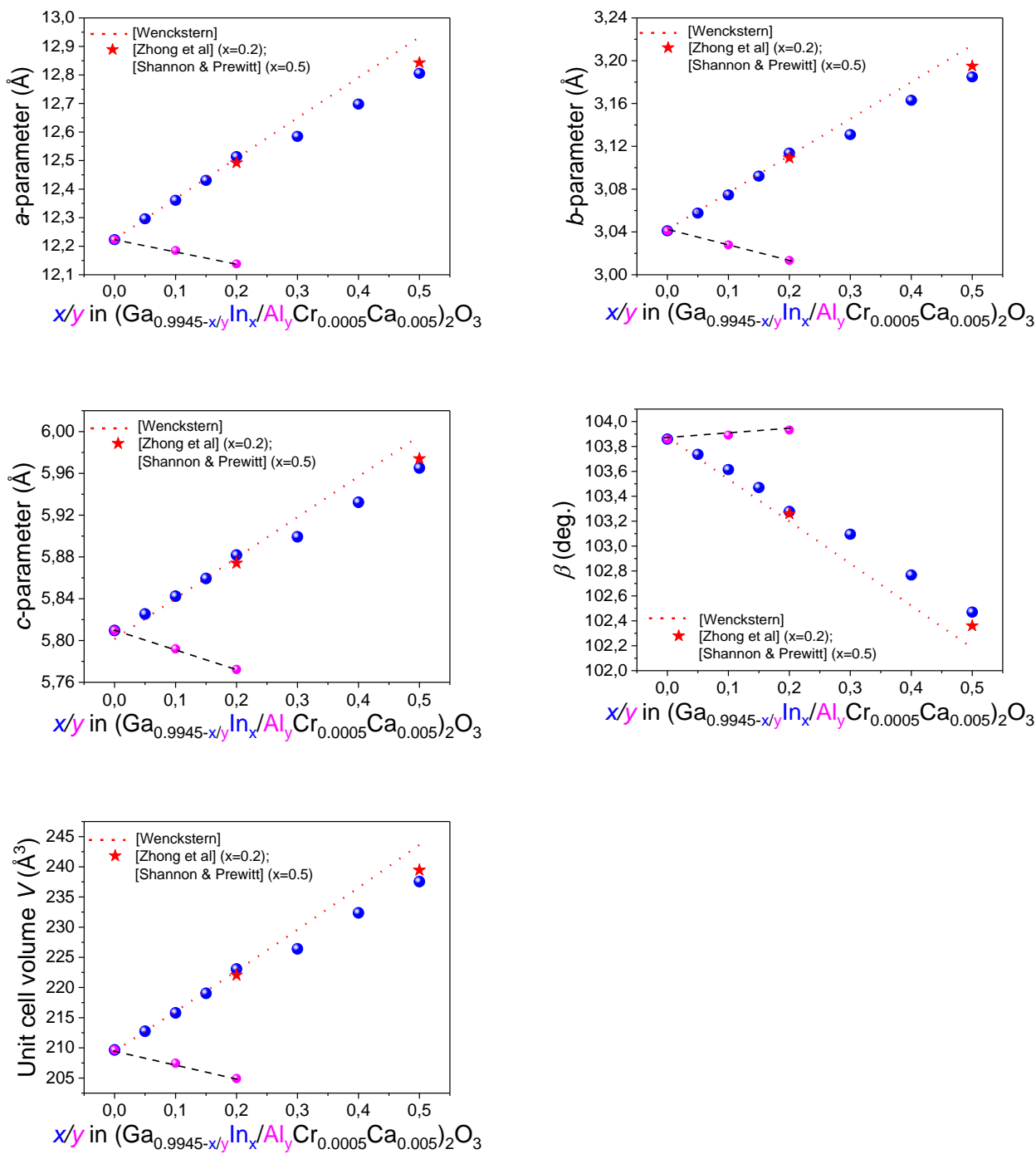

**Figure 4.** Evolution of the monoclinic lattice parameters and unit cell volume in the studied $(Ga_{1-x}In_x)_2O_3$:Cr, Ca and $(Ga_{1-y}Al_y)_2O_3$:Cr, Ca structures *vs* In and Al content (blue spheres and pink squares, respectively). Red stars correspond to the literature data for two In-containing materials from Refs. [38,40]. Red dotted lines represent linear fits performed by Wenckstem[39] for five data sets for the $(Ga_{1-x}In_x)_2O_3$ powder and single crystal samples taken from Refs. [38,50–53]

Regarding the pseudoternary Ga-Al-In oxides, the chosen synthesis method has proven inadequate for achieving the desired β-Ga$_2$O$_3$ structural type in samples with nominal host compositions of (Ga$_{0.25}$Al$_{0.5}$In$_{0.25}$)$_2$O$_3$ and (Ga$_{0.33}$Al$_{0.34}$In$_{0.33}$)$_2$O$_3$. XRD analysis of these materials (see Figure S1) did not reveal any matches within the Ga, Al, In and O containing phases listed in ICDD PDF-4+ and COD databases. This suggests that the monoclinic β-Ga$_2$O$_3$ structure cannot simultaneously accommodate such significant disparity in size Al$^{3+}$ and In$^{3+}$ cations, despite their average cation radius being close to that of Ga$^{3+}$. Consequently, the applied experimental conditions lead to the formation of new phase(s) with unknown structures within the Ga-Al-In-O quaternary system. Identifying these phases requires a comprehensive investigation of the phase behavior in the Ga$_2$O$_3$-Al$_2$O$_3$-In$_2$O$_3$ pseudoternary system, which is currently underway.

**PL and PLE of Cr$^{3+}$ in the β-(Ga-Al)$_2$O$_3$ and β-(Ga-In)$_2$O$_3$ alloys**

The photoluminescence (PL) and photoluminescence excitation (PLE) spectra of the studied compounds are shown in Figure 5a. The PLE spectra have two broad bands characteristic of Cr$^{3+}$ ions in octahedral coordination: the lower-energy band with a maximum at 605 nm arising from the $^4A_2 \rightarrow {}^4T_2$, and a subsequent band at 440 nm ($x = 0$) caused by the spin-allowed $^4A_2 \rightarrow {}^4T_1$ transitions. The third excitation band with a peak at about 300 nm ($x = 0$) is related most probably to the charge transfer (CT) transitions.[27,40,41] Additionally, above these excitation bands, an edge-like excitation caused by the band-to-band transitions (≤270 nm for $x = 0$) is observed. Upon partial substitution of Ga$^{3+}$ cations by smaller Al$^{3+}$ cations, all the excitation bands as well as the edge-like band exhibit a shift towards higher energies. Conversely, when Ga$^{3+}$ is replaced by larger In$^{3+}$ cations, a red shift is observed. This behavior is consistent with the results reported recently Refs.[34,40,41,44] and arises from the alteration in crystal field strength experienced by Cr$^{3+}$ ions due to the replacement of Ga by Al ions or In. It is worth highlighting that the main reason for the larger shift of the edge-like band and the CT band is alteration in bandgap. In contrast, the shift observed in the bands attributed to the $^4A_2 \rightarrow {}^4T_2$ and $^4A_2 \rightarrow {}^4T_1$ transitions are primarily a consequence of changes in the crystal field strength. The band-to-band excitation of Cr$^{3+}$ ions can be used as a measure of the optical bandgap of the studied materials. To illustrate this, Figure 6 showcases the optical bandgap (determined by locating the maximum of the band-to-band excitation band) as a function of the average cation (Al, Ga, In) radius. Based on the crystal structure results presented above indicating that In primarily replaces the octahedrally coordinated Ga, the bandgap evolution in Figure 6 is presented as a function of the average ionic radius of the octahedrally coordinated cations (CN=6). As it is seen from the figure, the bandgap value of the monoclinic (Ga-Al)$_2$O$_3$ and (Ga-In)$_2$O$_3$ oxides is gradually decreased when Al is replaced by Ga and then by In. This trend is in agreement with the previous experimental,[54] and theoretical[43] findings. A contrasting trend is evident in the unit cell volume, which demonstrates near linear growth with increasing ionic radii in both (Ga$_{1-x}$In$_x$)$_2$O$_3$:Cr, Ca and (Ga$_{1-y}$Al$_y$)$_2$O$_3$:Cr, Ca series as illustrated in Figure 7. These plots underscore a discernable correlation between the band gap of the Al- and In- substituted series and the unit cell volume of the β-Ga$_2$O$_3$-based materials.

Figure 5b shows the emission spectra of Cr$^{3+}$ ions in the studied compounds. For the samples with aluminum ($y$ = 0.1 and 0.2) and low indium content ($x \leq 0.1$), a sharp line at about 690 nm (the spin-forbidden $^2$E → $^4$A$_2$ transition) is present at room temperature alongside the broadband (the spin-allowed $^4$T$_2$ → $^4$A$_2$ transition) emission. With increasing indium content, the sharp emission line diminishes giving way to increasing dominance of the broadband emission. Besides, the width of the broad band increases, and their maximum progressively shifts towards NIR. This agrees with the recent findings reported by Zhong, J. et al.,[40] and stems from the decrease in crystal field strength when Ga is substituted by In. This vividly demonstrates how the *chemical pressure* alters the room-temperature emission of Cr$^{3+}$ in the studied compounds.

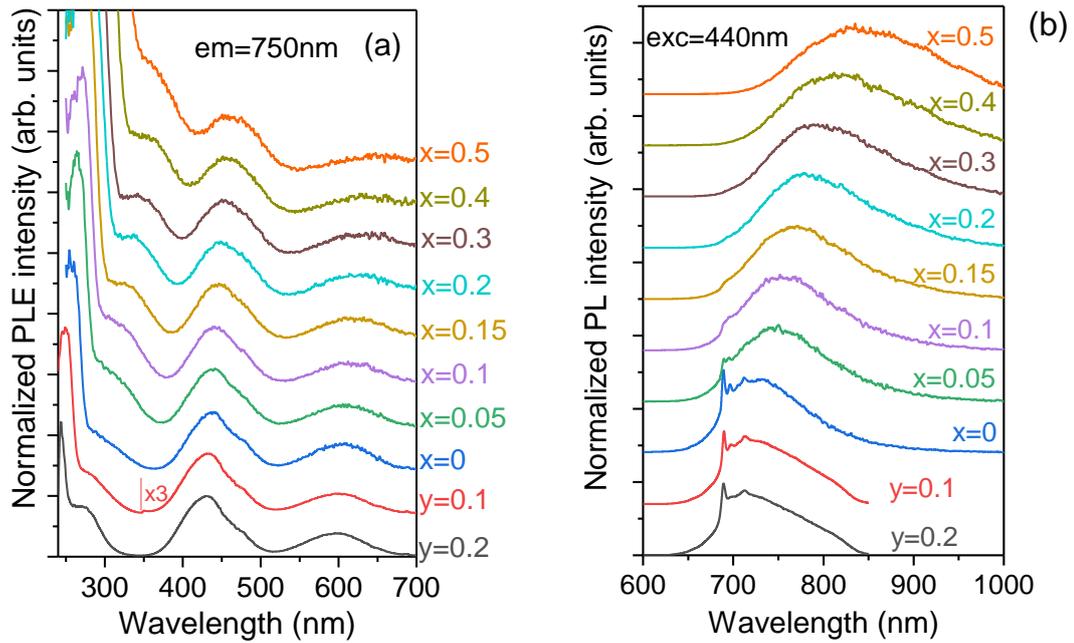

**Figure 5.** Normalized room-temperature PLE (a) and PL (b) spectra of $Cr^{3+}$ in β-$(Ga_{1-x}In_x)_2O_3$ ($x = 0…0.5$) and β-$(Ga_{1-y}Al_y)_2O_3$ ($y = 0.1, 0.2$) solid solutions calcinated at 1500°C.

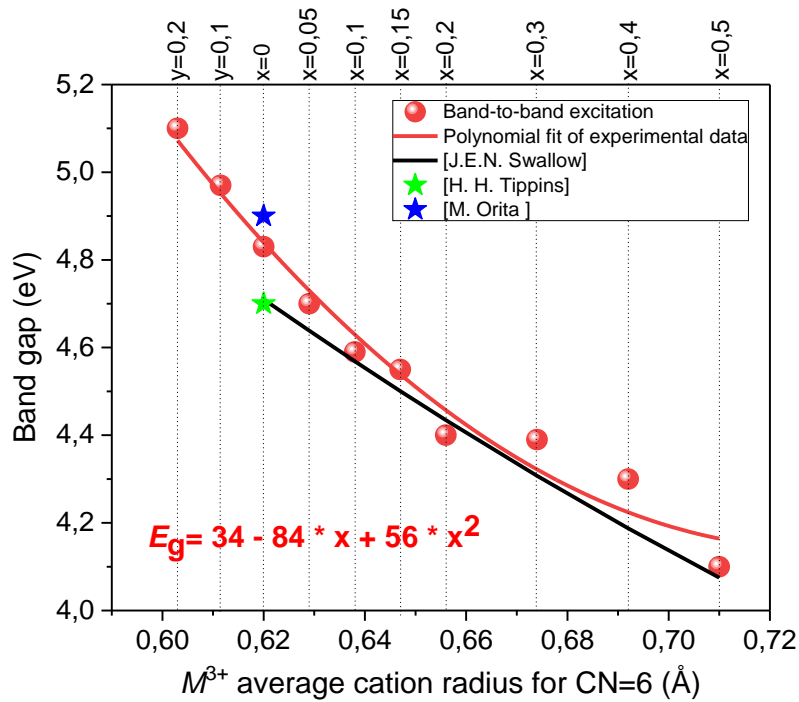

**Figure 6.** Optical bandgap values as a function of the $M^{3+}$ ($M$ = Al, Ga, In) average cation radius (coordination number, CN=6) for the monoclinic pseudobinary $(Ga-Al)_2O_3$ and $(Ga-In)_2O_3$ oxides (red spheres represent experimental data, red line – polynomial fit). Blue and green stars correspond to the bandgaps for β-$Ga_2O_3$ taken from Refs.[19,25] Black line represents fitting performed by Swallow et al.[54]

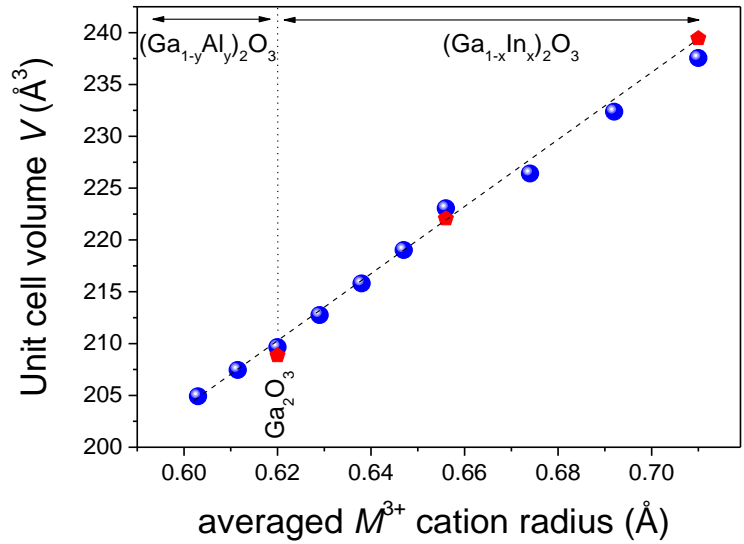

**Figure 7.** Unit cell volume as a function of the $M^{3+}$ ($M$ = Al, Ga, In) average cation radius (coordination number, CN=6) for the monoclinic $M_2O_3$ oxides.

*Effect of calcination temperature.* It was confirmed by us experimentally that, after the heat treatment at 1500°C, the $Cr^{3+}$ emission efficiency of the $(Ga_{1-x}In_x)_2O_3$ samples increased by almost an order of magnitude with respect to the samples calcined at 1200°C.

Figure 8a demonstrates the temperature-dependent evolution of the $Cr^{3+}$ emission spectrum using the $x = 0.1$ composition of the β-$(Ga_{1-x}In_x)_2O_3$ solid solution as an example. It is evident that the relative intensity of the narrow-line emission at about 690 nm in relation to the broad-band emission decreases as temperature increases, so that the narrow line is barely noticeable at room temperature. This temperature behavior indicates the thermalization effect between the closely situated $^2E$ and $^4T_2$ emitting levels, which aligns with the prior findings for pure β-$Ga_2O_3$:$Cr^{3+}$.[27] However, the low-temperature PLE spectrum (see Figure 8b) exhibits discrepancies between the narrow-line and the broad-band emissions. This suggests the potential existence of at least two different types of $Cr^{3+}$ emitting centers in such solid solutions.

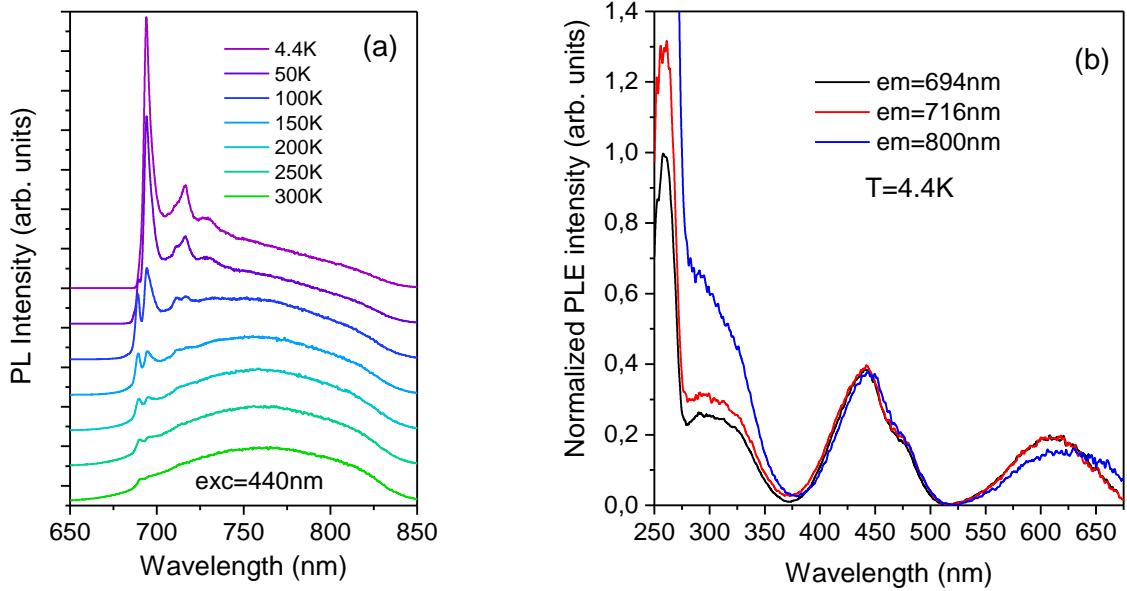

**Figure 8.** Temperature-dependent PL (a) and PLE (b) spectra of $Cr^{3+}$ emission in $\beta$-$(Ga_{0.9}In_{0.1})_2O_3$.

Indeed, the high-resolution low-temperature spectra measured in the region of the zero-phonon line (ZPL) of $Cr^{3+}$ emission for the studied $\beta$-(Ga-Al)$_2$O$_3$ and $\beta$-(Ga-In)$_2$O$_3$ solid solutions, shown in Figure 9, confirm a complex emission structure. In particular, it was revealed that the $R_1$-line emission comprises two closely located lines for all the studied solid solutions except for the pure $\beta$-Ga$_2$O$_3$:Cr and $\beta$-(Ga$_{0.5}$In$_{0.5}$)$_2$O$_3$:Cr samples. It should be noted that the observed splitting demonstrated in Figure 9 is not related to the $R_2$-line, *i.e.* splitting of the $^2$E level caused by a decrease in symmetry of the oxygen octahedra. As illustrated in Figure 10, the $R_2$-line appears somewhat displaced with the increase in temperature. Moreover, Figure 10 unequivocally demonstrates that both the $R_1$ and $R_2$ lines in this particular composition are split into two components. Hence, the red, green, and blue lines in Figure 9 represent the $R_1$ ZPL emission line of three types of $Cr^{3+}$ centers in the studied $\beta$-(Ga-Al)$_2$O$_3$ and $\beta$-(Ga-In)$_2$O$_3$ solid solutions. A superposition of the "red" and "green" peaks is observed in the Ga-Al oxides, while the "green" and "blue" peaks are characteristic of the Ga-In oxides. Notably, for pure $\beta$-Ga$_2$O$_3$:Cr, only one type of $R_1$-line (denoted "green" in our notation) is observed. Conversely, another type of $R_1$-line (referred to as "blue") is predominantly observed for $\beta$-(Ga$_{0.5}$In$_{0.5}$)$_2$O$_3$:Cr.

It is intriguing to observe the variation in the "red", "green" and "blue" ZPLs as the composition of the host lattice undergoes alteration. Evidently, an increase in the Al content, leads to an augmented relative intensity of the "red" line, whereas higher In content gives rise to the emergence of the "blue" line. The half-width of each line increases with the rising intensity of the complementary component. Consequently, the "green" and "blue" lines are at their narrowest when they occur individually. Simultaneously, there is a discernable shift of each line with a change in composition. As the In content increases, the "blue" line shifts towards longer wavelengths, while the "green" line shifts towards shorter wavelengths (corresponding to higher energies). This trend implies that the $Cr^{3+}$ centers responsible for the "green" ZPL experience a diminished *nephelauxetic effect* signifying a reduction of the covalency of chemical bonding between these $Cr^{3+}$ centers and the host ligands[55] as In content increases. Conversely, the centers accountable for the "blue" ZPL exhibit increase of covalency of bonding as In content rises.

Given these observations for the multiple ZPL of $Cr^{3+}$ centers observed in the investigated β-(Ga-Al)$_2$O$_3$ and β-(Ga-In)$_2$O$_3$ solid solutions, it can be deduced that the "red", "green" and "blue" ZPLs belong to $Cr^{3+}$ centers occupying the octahedral sites typical for Al, Ga and In, respectively. The crystal structure findings as well as the spectroscopic results of $Cr^{3+}$ ZPL emission presented above unequivocally indicate that these Al-, Ga- and In-centered octahedra in the alloyed host lattices undergo systematic alteration compared to "pure" compounds. However, complete averaging of local distances between cations and oxygen anions does not occur, and the $Cr^{3+}$ dopant replacing some of these octahedrally-coordinated cations can be used to discern whether the octahedron is primarily Al-, Ga- or In-centered.

The spectroscopic results on $Cr^{3+}$ doping presented in Figure 9 provide further insight into the mechanism of substitution of Ga by In in the β-(Ga-In)$_2$O$_3$ solid solutions. The gradual replacement of the "green" ZPL of $Cr^{3+}$ by the "blue" line as Ga is substituted by In, along with the exclusive presence of the "blue" line in β-(Ga$_{0.5}$In$_{0.5}$)$_2$O$_3$, confirms that In ions exclusively replace the octahedrally-coordinated Ga. Consequently, in the β-(Ga$_{0.5}$In$_{0.5}$)$_2$O$_3$:Cr compound no $Cr^{3+}$ centers remain in the octahedrally-coordinated Ga sites. This observation fully agrees with the crystal structure results presented above.

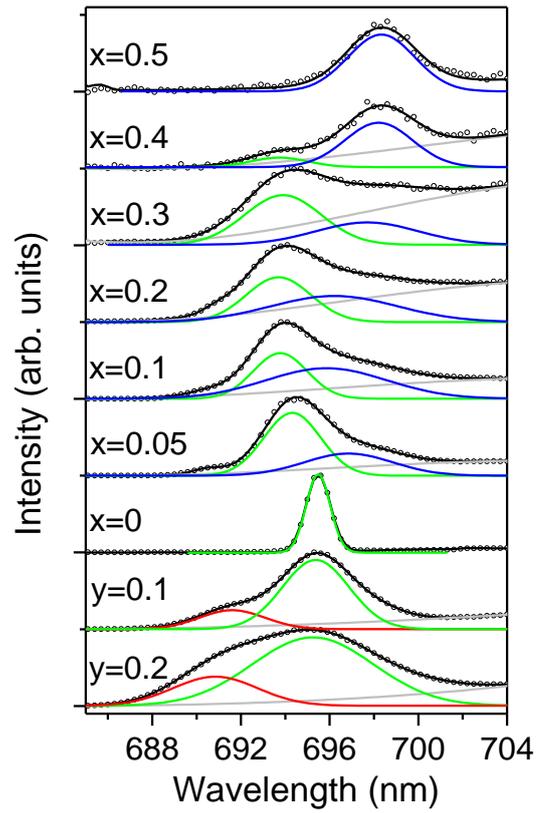

**Figure 9.** High-resolution $Cr^{3+}$ ZPL ($R_1$-lines) emission spectra for the studied $\beta$-$(Ga_{1-x}In_x)_2O_3$ and $\beta$-$(Ga_{1-y}Al_y)_2O_3$ solid solutions measured at $T$=4.4K (open circles) and their fitting with a Gaussian multipeak function (solid lines).

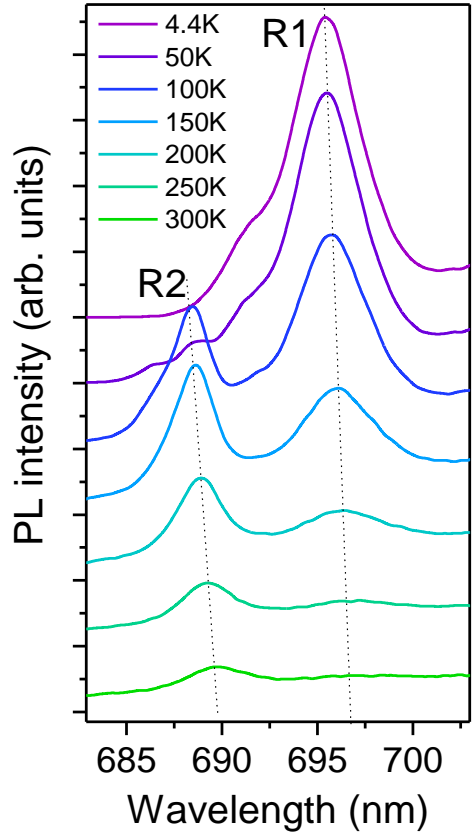

**Figure 10.** Temperature evolution of $Cr^{3+}$ ZPL emission spectra in $\beta$-$(Ga_{0.9}Al_{0.1})_2O_3$ sample.

**PL decay and thermometric performance of the $Cr^{3+}$-doped $\beta$-(Ga-Al)$_2$O$_3$ and $\beta$-(Ga-In)$_2$O$_3$ alloys**

The photoluminescence decay kinetics were revealed to be non-single exponential for all the studied compounds except the pure $\beta$-Ga$_2$O$_3$:Cr ($x$=0) sample, the decay kinetics for which were published previously.[30] As an illustration, Figure 11a demonstrates the decay kinetics separately recorded for the "green" and "blue" ZPL (refer to Figure 9) of $Cr^{3+}$ emission for the $x$=0.2 sample. Fitting these decay kinetics using two-exponential decay formula:

$$y = y_0 + A_1 e^{-x/\tau 1} + A_2 e^{-x/\tau 2} \qquad (1)$$

as shown in the Figure 11a, yield similar values of the characteristic decay time of $\tau_1$=0.3±0.05 ms and $\tau_2$=1.4±0.1 ms with the intensity ratios $A_1/A_2$~1.4 and $A_1/A_2$~0.4 for the "green" and "blue" components, respectively. Juxtaposing these ratios with the data shown in Figure 9 we infer that the shorter decay time $\tau_1$ corresponds to the "green" component, while the longer decay time $\tau_2$ is related to the "blue" component for this sample. These findings also affirm that the non-single exponential PL decays of the studied alloyed β-(Al-Ga)$_2$O$_3$:Cr and β-(Ga-In)$_2$O$_3$:Cr compounds are caused by the presence of multiple (at least of two types) Cr$^{3+}$ centers within each of these alloyed compounds.

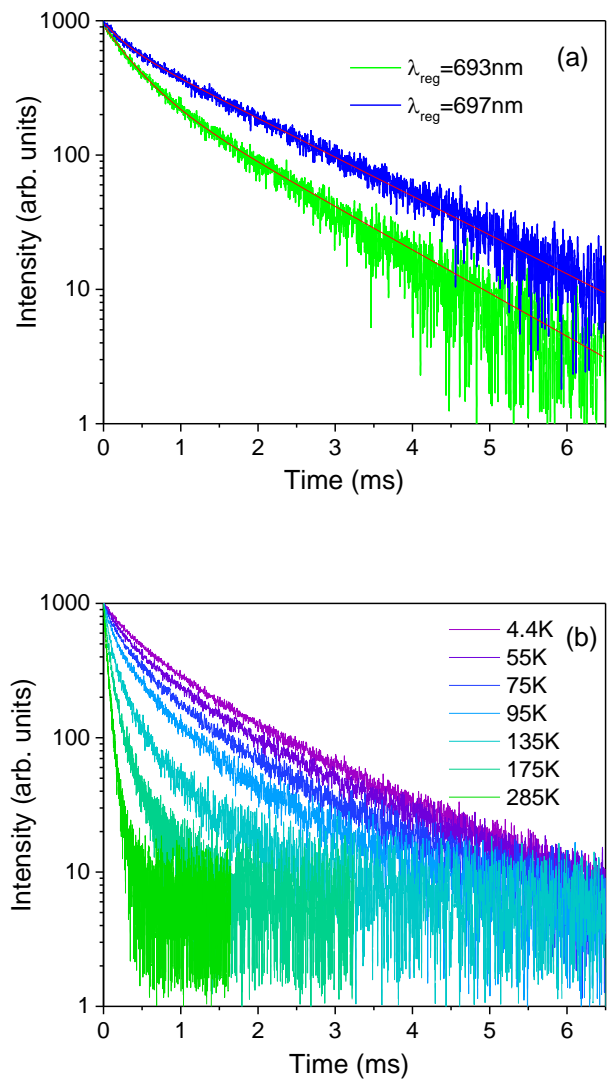

**Figure 11.** The PL decay curves of Cr$^{3+}$ ZPL emission in the β-(Ga$_{0.8}$In$_{0.2}$)$_2$O$_3$ sample recorded separately for the "green" ($\lambda_{reg}$=693 nm, bandpass = 1 nm) and "blue" ($\lambda_{reg}$=697 nm, bandpass = 1 nm) components at $T$ = 4.4 K (a). The solid red lines in (a) represent fits using a two-exponential decay equation with $\tau_1$=0.3±0.05 ms and $\tau_2$=1.4±0.1 ms (see text for details). Temperature evolution of the PL decay of Cr$^{3+}$ ZPL emission in the β-(Ga$_{0.8}$In$_{0.2}$)$_2$O$_3$ sample registered overally ($\lambda_{reg}$=694 nm, bandpass = 3 nm) in the vicinity of ZPL emission (b).

Temperature evolution of the PL decay of $Cr^{3+}$ emission in the $\beta\text{-}(Ga_{0.8}In_{0.2})_2O_3$ sample is shown in Figure 11b. As is seen from the figure, the decay kinetics exhibit dramatical shortening as the temperature rises. Due to the non-single exponential shape of the decay kinetics, to evaluate the lifetimes, we employed the mean weighted decay time:

$$\bar{\tau} = \frac{\int_0^{t_{max}} t \cdot I(t)\, dt}{\int_0^{t_{max}} I(t)\, dt}, \quad (2)$$

where 0 and $t_{max}$ represent the initial and final time points of the measurement range, accordingly, while $I(t)$ denotes the luminescence intensity at time $t$, similarly to the approach used by Zhong et al.[40]

The mean decay times of $Cr^{3+}$ emission obtained using this approach are plotted against temperature for the investigated $\beta\text{-}(Ga_{1-x}In_x)_2O_3$ and $\beta\text{-}(Ga_{1-y}Al_y)_2O_3$ solid solutions in Figure 12. The acquired temperature dependances $\tau = f(T)$ demonstrate a gradual reduction of decay time constant as the temperature rises. This behavior aligns with a well-documented pattern observed in the luminescence of various materials wherein the increased likelihood of radiative transition is attributed to heightened non-radiative de-excitation of emitting levels. To comprehensively address this behavior, we adopted a model of the temperature dependence of the luminescence decay time in the materials activated by the transition metals model formulated and validated by Mykhaylyk et al.[30]

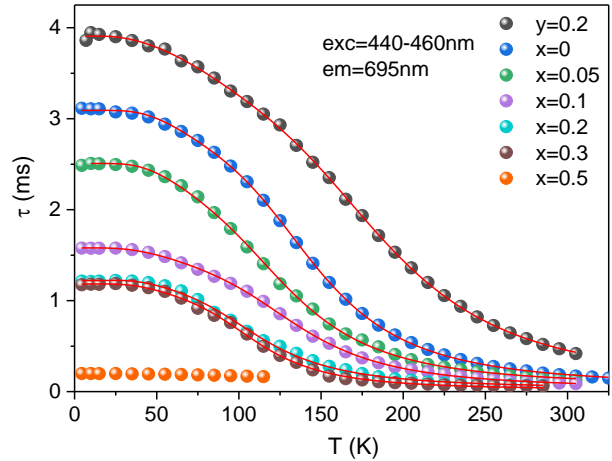

**Figure 12.** Temperature dependencies of the mean decay times of $Cr^{3+}$ emission (registered in ZPL) in the studied $\beta\text{-}(Ga_{1-x}In_x)_2O_3$ and $\beta\text{-}(Ga_{1-y}Al_y)_2O_3$ solid solutions of different compositions. The red lines represent fits of the experimental data using Eq. (3) with parameters summarized in Table 5.

The model takes into account variations in the population of emitting states due to the combined effect of thermally induced depopulation and phonon-assisted relaxation of the emission center. In the framework of this model, the radiative decay rate $1/\tau(T)$ is given by the following expression:

$$\tau(T) = \frac{1+exp\left(-\frac{D}{kT}\right)+6exp\left(-\frac{\Delta E}{kT}\right)}{\frac{1}{\tau_1}coth\left(\frac{E_p}{2kT}\right)+\frac{1}{\tau_2}coth\left(\frac{E_p}{2kT}\right)exp\left(-\frac{D}{2kT}\right)+\frac{6}{\tau_3}exp\left(-\frac{\Delta E}{kT}\right)} \quad . \quad (3)$$

In this equation $1/\tau_i$ ($i$= 1, 2 and 3) are the radiative decay rates of the $\bar{E}$, $\overline{2A}$, and $^4T_2$ levels respectively. $D$ and $\Delta E$ are the energy differences between the $\bar{E}$, $\overline{2A}$, and $^4T_2$ levels respectively, $E_p$ stands for "effective energy" of the phonons responsible for the exchange with the sidebands, $k$ is the Boltzmann constant and T is the absolute temperature.

Upon application of Eq. (3) perfect agreement of experimental data and model was achieved across the entire temperature range as clearly illustrated by the solid lines in Figure 12. The parameters of the fit are summarised in Table 3. Upon inspection of the fitting parameters, several discernable trends emerge. Foremost among them is the evident decrease in activation energy $\Delta E$ with an increase of indium content, signifying a reduction in the energy barrier for the non-radiative depopulation of emitting levels. This effect is concomitant with a decrease in the decay time constant τ3, which is inversely related to the probability of radiative transitions from $^4T_2$ level and subsequently influences the temperature-dependent dynamics. The next noteworthy feature pertains to the relationship between the radiative decay constants of $\tau_1$ and $\tau_2$. Their ratio plays a pivotal role in determining the radiative exchange between the $\bar{E}$ and $\overline{2A}$ levels of $Cr^{3+}$. In numerous Cr-doped oxides, this ratio falls below unity because at very low temperatures the upper level with longer radiative lifetime is populated by phonon-mediated transitions from the lower emission level of $Cr^{3+}$.[30] This leads to a decrease in the measured decay time constant. However, this does not hold for gallium oxide and its alloys, where the lower level ($\bar{E}$) exhibits a longer radiative lifetime and hence where $\tau_1/\tau_2$>1. (It is worth noting that, for an as yet unknown reason, $(Ga_{0.8}In_{0.2})_2O_3$ deviates from this pattern.) As a result, heating leads to the thermal population of the upper $\overline{2A}$ level and a consequent steady augmentation of the overall transition rate. This is manifested as a consistent decline in the measured decay time constant with increasing temperature in Cr-doped gallium-indium alloys. Finally, the reduction in effective phonon energy $E_p$ with increasing indium concentration finds its rationale in the greater mass of In relative to Ga, which leads to a decrease in the vibrational frequencies of the Ga(In)-O bonds.

**Table 3.** Parameters of fits obtained from the temperature dependence of the PL decay time Eq.(3) (5) for $Cr^{3+}$ emission in β-$(Ga_{0.8}Al_{0.2})_2O_3$ and β-$(Ga_{1-x}In_x)_2O_3$ ($x$ = 0…0.5) and.

| Sample composition ($x$ value) | $\tau_1$, ms | $\tau_2$, ms | $E_p$, meV | $D$, meV | $\tau_3$, μs | $\Delta E_1$, meV |
|---|---|---|---|---|---|---|
| $(Ga_{0.8}Al_{0.2})_2O_3$ | 3.91±0.01 | 2.83±0.10 | 25.7±0.8 | 11 | 36.7±2.8 | 104.8±1.7 |
| $Ga_2O_3$ | 3.09±0.01 | 1.05±0.04 | 34.9±3.3 | 18 | 30.4±1. | 82.9±1.0 |
| $(Ga_{0.95}In_{0.05})_2O_3$ | 2.51±0.01 | 1.22±0.5 | 20.1±0.4 | 16 | 33.4±2.7 | 72.9±1.3 |
| $(Ga_{0.9}In_{0.1})_2O_3$ | 1.58±0.01 | 1.17±0.05 | 20.1±0.1 | 10 | 17.7±0.9 | 75.2±0.9 |
| $(Ga_{0.8}In_{0.2})_2O_3$ | 1.22±0.01 | 2.00±0.6 | 13.4±0.9 | 9 | 18.9±2.2 | 60.0±1.8 |
| $(Ga_{0.7}In_{0.3})_2O_3$ | 1.18±0.01 | 1.15±0.3 | 15.1±1.2 | 12 | 9.8±1.6 | 64.5±3.2 |

* The value of $D$ is fixed to be equal to the energy splitting of the $^2E$ level

The assessment of the studied materials suitability for thermometry can be derived from Figure 13, illustrating the temperature-dependent variations of specific sensitivity $|\Delta\tau/\Delta T|\tau^{-1}$. In this context, the specific sensitivity of about 1.7 %/K at 170 K observed in β-$Ga_2O_3$:Cr agrees with prior findings.[30] As it is seen from Figure 13, alloying β-$Ga_2O_3$ with $Al_2O_3$ leads to a reduction in specific sensitivity accompanied by a shift of their maximum towards higher temperatures (1.25 %/K at 220 K for the β-$(Ga_{0.8}Al_{0.2})_2O_3$ sample). Conversely, when β-$Ga_2O_3$ is alloyed with $In_2O_3$, the maximum of the specific sensitivity progressively shifts towards lower temperatures. Notably, the $(Ga_{0.7}In_{0.3})_2O_3$ exhibits the highest specific sensitivity reaching approximately 2.35 %/K at 130 K.

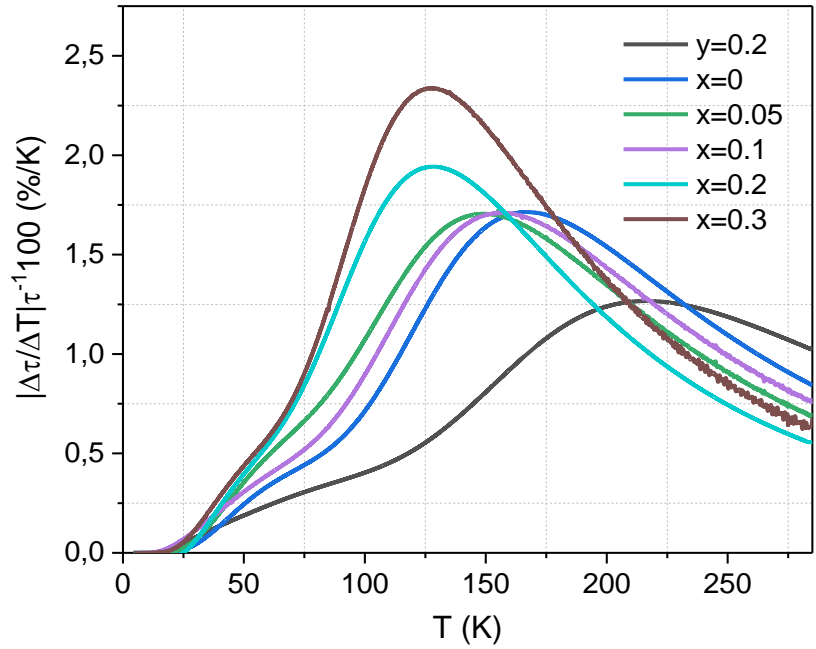

**Figure 13.** Specific sensitivity of the decay time luminescence thermometers based on the studied $Cr^{3+}$-doped β-$(Ga_{1-x}In_x)_2O_3$ and β-$(Ga_{1-y}Al_y)_2O_3$ solid solutions. Solid lines are derived from the fits using Eq. (3).

# Thermoluminescence and persistent luminescence of the Cr and Ca co-doped β-(Ga-Al)₂O₃ and β-(Ga-In)₂O₃ alloys

After exposure to UV (we used 300 nm), all examined Cr and Ca co-doped β-(Ga$_{1-x}$In$_x$)$_2$O$_3$ and β-(Ga$_{1-y}$Al$_y$)$_2$O$_3$ solid solutions reveal a quite intensive thermal glow in the red-NIR spectral range corresponding to the Cr$^{3+}$ emission of samples (see Figure 14). It should be noted that the thermal glow, as well as an afterglow (Figure 15), of studied materials is observed only following UV excitation of CT band (see Figure 5) and is not induced by the exposure to visible light into the $^4A_2 \rightarrow {^4T_2}$ or $^4A_2 \rightarrow {^4T_1}$ intracenter excitation bands of Cr$^{3+}$.

The thermal glow curve for the β-Ga₂O₃:Cr, Ca ($x$=0) sample manifests a dual peak pattern with maxima at about 320 and 365 K (at linear heating with rate 1 K/s) and a less-prominent peak at higher temperatures around 500 K. Such structure of thermal glow is typical for Cr-doped as well as Cr and Ca co-doped β-Ga₂O₃. Similar peaks observed in β-Ga₂O₃:Cr, Mg crystals, and ceramics have been attributed to deep trap levels related to the Cr$^{3+}$ close to oxygen vacancies and oxygen vacancies with trapped electrons.[45]

As one can see from Figure 14, after alloying with Al₂O₃, the primary TL peak shifts towards a higher temperature, exhibiting maximum at ca. 455 K for the $y$=0.2 sample. Conversely, in samples alloyed with In₂O₃, the peak gradually moves towards lower temperatures with increasing In content. The characteristic shape of this TL peak in all Al- or In-alloyed samples, particularly the prolongated high-temperature tail in comparison with the swift rising part of the peak, suggests that this TL peak remains composite (consisting of two or possibly more overlapping peaks) similar to pristine β-Ga₂O₃. The fact that the structure of the thermal glow remains generally the same with only changes in the position of the peak maxima implies that we are dealing with the same type of traps, albeit with different depths as a consequence of changes in the bandgap of the host lattice. Notably, with increasing In content, the intensity of the main TL peak (which is slightly above room temperature for $x$=0) decreases in comparison to the higher-temperature peak, leading to the dominance of the latter, which is already slightly above room temperature for $x$=0.5. Furthermore, alloying with Al₂O₃ or In₂O₃ introduces a significant broadening of the main TL peak compared to pristine β-Ga₂O₃.

The intensity of the afterglow correlates with the position of the TL peak at or slightly above room temperature. As it is seen from Figure 15, among all the examined samples the largest afterglow intensity during the first 100 s was found for the $x$=0.05 sample, owing to its

TL peak maximum being closest to room temperature. It is worthwhile noting that the sample with the highest In content ($x=0.5$) also demonstrates some afterglow primarily attributable to the higher-temperature TL peak, which is already close proximity to room temperature.

In such a way, the results presented in this section clearly demonstrate how the *bandgap engineering* via changing the chemical composition of the host lattice offers an effective means of fine-tuning the thermoluminescent and persistent luminescent properties of the studied materials.

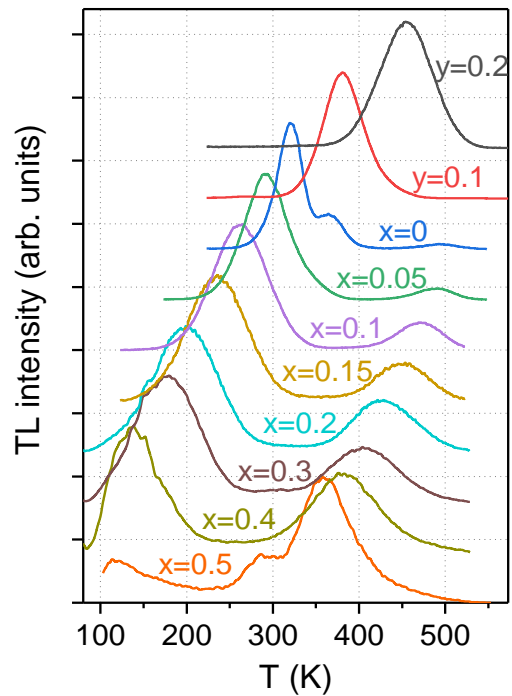

**Figure 14**. Normalized thermal glow registered at 750 nm of the studied Cr and Ca co-doped β-$(Ga_{1-x}In_x)_2O_3$ and β-$(Ga_{1-y}Al_y)_2O_3$ solid solutions after exposure to 300 nm UV radiation, heating rate 1 K/s.

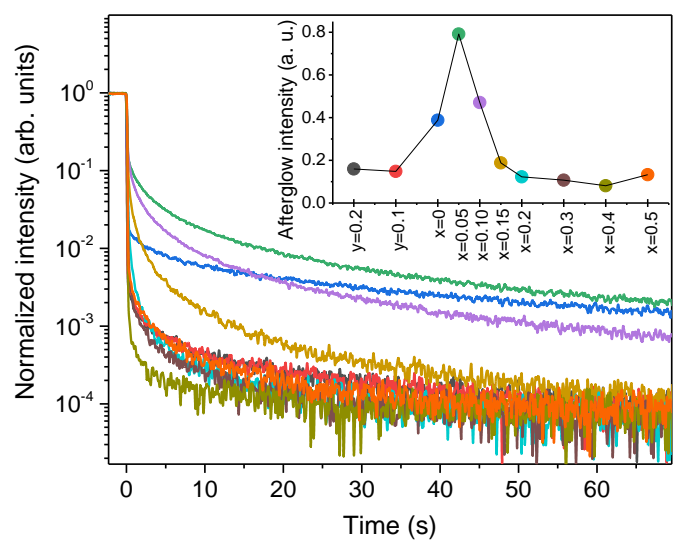

**Figure 15.** Afterglow decay kinetics of the studied Cr and Ca co-doped β-$(Ga_{1-x}In_x)_2O_3$ and β-$(Ga_{1-y}Al_y)_2O_3$ solid solutions monitored at 750 nm after exposure to UV (300 nm). The inset represents the intensity of the glow integrated during the first 100 s.

**CONCLUSIONS**

All pseudobinary compounds within the $(Ga_{1-x}In_x)_2O_3$ ($x$ =0; 0.05; 0.1; 0.15; 0.2; 0.3; 0.4; 0.5) and $(Ga_{1-y}Al_y)_2O_3$ ($y$ = 0.1; 0.2) series obtained by solution combustion method were identified as single-phase structures of β-$Ga_2O_3$-type. These compounds were used to study photoluminescent, thermoluminescent and persistent luminescent properties of $Cr^{3+}$ ions. The monoclinic crystal lattice parameters *a*, *b*, *c,* and the unit cell volume of the studied pseudobinary compounds increase systematically with increasing average cation radii in $(Ga_{1-x}In_x)_2O_3$ and $(Ga_{1-y}Al_y)_2O_3$ series, whereas the monoclinic angle β decreases. The refinement of the site occupancies showed that $Al^{3+}$ ions enter both tetrahedral and octahedral positions of β-$Ga_2O_3$ structure but with a clear preference for the octahedral position. In contrast, $In^{3+}$ ions were primarily found in the octahedral positions, whereas the tetrahedral sites remain occupied by $Ga^{3+}$ ions. Consequently, the sample of β-$(Ga_{0.5}In_{0.5})_2O_3$ is predominantly composed of Ga tetrahedra and In octahedra.

The obtained pseudoternary compounds $(Ga_{0.25}Al_{0.5}In_{0.25})_2O_3$ and $(Ga_{0.33}Al_{0.34}In_{0.33})_2O_3$ were found to possess the same, yet unidentified crystal structure, which is different from the monoclinic structure of β-$Ga_2O_3$. This suggests that the monoclinic structure does not tolerate such amounts of significantly different in size cations like $Al^{3+}$ and $In^{3+}$, despite the fact that the average cation radius here is close to $Ga^{3+}$.

The optical bandgap of the investigated $(Ga-Al)_2O_3$ and $(Ga-In)_2O_3$ monoclinic compounds estimated from the $Cr^{3+}$ photoluminescence excitation spectra progressively decreases as Al is replaced by Ga and then subsequently by In. This finding aligns with earlier individual results reported for pure and In-containing $\beta$-$Ga_2O_3$.

Our photoluminescence studies provide compelling evidence for a diminishing crystal field strength experienced by $Cr^{3+}$ ions when Al/Ga and Ga/In ratios in the studied pseudobinary compounds decrease. Moreover, through low-temperature studies of photoluminescence and decay time in the Cr-doped $(Ga-Al)_2O_3$ and $(Ga-In)_2O_3$ compounds we unveiled the existence of multiple $Cr^{3+}$ centers, which were attributed to the Al-, Ga- and In-centered octahedra in the studied host lattices. This suggests that a complete averaging of local distances between cations and oxygen anions in the studied alloys does not occur, and the $Cr^{3+}$ dopant substituting these octahedrally-coordinated cations emerges a discerning probe enabling identify whether octahedron is primarily Al-, Ga- or In-centered.

Our studies confirm that $Ca^{2+}$ co-doping enhances thermoluminescence of the studied materials through the generation of defects associated with oxygen vacancies. The obtained results clearly demonstrate that bandgap engineering via alteration of the chemical composition of the host lattice allows to tune the thermoluminescent and persistent luminescent properties of the studied materials by changing the energy depth of the traps responsible for the thermoluminescence. The alteration of position and broadening of TL peak for mixed compositions represents a potential strategy for tailoring the NIR persistent properties of the $\beta$-$Ga_2O_3$:Cr-based phosphors.

Finally, we demonstrated the capability to fine-tuning the thermometric performance of the studied phosphors by manipulating the chemical composition of the host lattice. This adjustment allows for control of temperature range and the maximal specific sensitivity of the decay time luminescence thermometers based on the studied $Cr^{3+}$-doped $(Ga-Al)_2O_3$ and $(Ga-In)_2O_3$ solid solutions. These findings highlight the significant potential of the studied materials for application in cryogenic luminescence thermometry. In particular, alloying of $\beta$-$Ga_2O_3$ with 30% of $In_2O_3$ leads to an almost twofold increase in the specific sensitivity of $\beta$-$Ga_2O_3$:$Cr^{3+}$ while simultaneously lowering the temperature of maximal sensitivity from 170 to 130 K.

# SUPPORTING INFORMATION

XRD data for the pseudoternary samples of nominal compositions $(Ga_{0.25}Al_{0.5}In_{0.25})_2O_3$ and $(Ga_{0.33}Al_{0.34}In_{0.33})_2O_3$.

## AUTHOR INFORMATION

**Corresponding Author**

**Vasyl Stasiv** – *Institute of Physics, Polish Academy of Sciences, Warsaw 02-668, Poland;* https://orcid.org/0000-0001-5477-8334; Email: stasiv@ifpan.edu.pl

**Authors**

- **Yaroslav Zhydachevskyy** – *Institute of Physics, Polish Academy of Sciences, Warsaw 02-668, Poland;* https://orcid.org/0000-0003-4774-5977; *Berdyansk State Pedagogical University, Shmidta Str. 4, Berdiansk 71100, Ukraine*
- **Vitalii Stadnik** – *Lviv Polytechnic National University, S. Bandera Str. 12, Lviv 79013, Ukraine*
- **Vasyl Hreb** – *Lviv Polytechnic National University, S. Bandera Str. 12, Lviv 79013, Ukraine*
- **Vitaliy Mykhaylyk** – *Diamond Light Source, Didcot, OX11 0DE, UK*
- **Leonid Vasylechko** – *Lviv Polytechnic National University, Lviv 79013, Ukraine*
- **Andriy Luchechko** – *Ivan Franko National University of Lviv, Lviv 79017, Ukraine*
- **Tomasz Wojciechowski** – *Institute of Physics, Polish Academy of Sciences, Warsaw 02-668, Poland*
- **Piotr Sybilski** – *Institute of Physics, Polish Academy of Sciences, Warsaw 02-668, Poland*
- **Andrzej Suchocki** – *Institute of Physics, Polish Academy of Sciences, Warsaw 02-668, Poland*

**Notes**

The authors declare no competing financial interest.

## ACKNOWLEDGMENTS

The work was supported by the Polish National Science Centre (project nos. 2018/31/B/ST8/00774 and 2021/40/Q/ST5/00336) and by the National Research Foundation of Ukraine (grant no. 2020.02/0373 "Crystalline phosphors' engineering for biomedical applications, energy saving lighting and contactless thermometry").

**TOC Graphic**